\documentclass{aastex63}

\newcommand{\msun}{\thinspace M_\odot}

\def\lesssim{\mathrel{\hbox{\rlap{\hbox{\lower4pt\hbox{$\sim$}}}\hbox{$<$}}}}
\def\gtrsim{\mathrel{\hbox{\rlap{\hbox{\lower4pt\hbox{$\sim$}}}\hbox{$>$}}}}
\newcommand{\cm}{\,{\rm cm}^{-3} } 
\newcommand{\km}{\,{\rm km\, s}^{-1}}

\shorttitle{Twin Protostellar Jets and Close Binary Formation}
\shortauthors{Saiki and Machida.}

\begin{document}

\title{Twin Jets and Close Binary Formation}

\correspondingauthor{Masahiro N. Machida}
\email{machida.masahiro.018@m.kyushu-u.ac.jp}

\author{Yu Saiki}
\author{Masahiro N. Machida}
\affil{Department of Earth and Planetary Sciences, Faculty of Sciences, Kyushu University, Fukuoka 819-0395, Japan}

\begin{abstract}
The formation of a close binary system is investigated using a three-dimensional {\rm resistive} magnetohydrodynamics simulation. 
Starting from a prestellar cloud, the cloud evolution is calculated until $\sim 400$\,yr after protostar formation. 
Fragmentation occurs in the gravitationally collapsing cloud and two fragments evolve into protostars.
The protostars orbit each other and a protobinary system appears.
A wide-angle low-velocity outflow emerges from the circumbinary streams that encloses two protostars, while each protostar episodically drives high-velocity jets. 
Thus, the two high-velocity jets are surrounded by the low-velocity circumbinary outflow. 
The speed of the jets exceeds $\gtrsim 100\km$. 
Although the jets have a collimated structure, they are swung back on the small scale and are tangled at the large scale due to the binary orbital motion.
A circumstellar disk also appears around each protostar.
%% and is embedded in the circumbinary disk.
In the early main accretion phase, the binary orbit is complicated, while the binary separation is within $<30$\,au.
For the first time, all the characteristics of protobinary systems recently observed with large telescopes are reproduced in a numerical simulation.
\end{abstract}


\section{Introduction} \label{sec:intro}
A large fraction of main-sequence stars are members of binary systems, and more than half of the stars exceeding $ 1\msun$ are binaries \citep{duquennoy91,moe17}.
The binary fraction of very young stars in star-forming regions is higher than that of field stars, which indicates that  many stars are born as binaries \citep{chen13,duchene13}.
Therefore, studies into star formation should consider binary formation and not focus only on single-star formation, in order to comprehensively understand the star formation process. 
Especially, the formation of close binaries with a separation of $\lesssim 10$\,au is important for clarifying the origin of gravitational waves, Type Ia supernovae, X-ray binaries and carbon enhanced metal poor stars \citep{riess98,remillard06,hansen16,abbott16}.
However, no convincing scenario has yet been proposed for close binary formation \citep{bodenheimer11}. 
	Past theoretical studies have proposed binary formation scenarios such as disk fragmentation, fission and capture \citep{tohline02}
\footnote{
Here, we commented on the formation of general binary systems but do not focus only on close binary systems. 
}, 
while observations have provided useful clues for understanding binary formation. 
Recent observations have unveiled newborn binary systems,
where observed circumbinary disks, circumbinary outflows and protostellar jets are clear evidence of mass accretion occurring in the forming binary systems \citep{itoh00,itoh01,hioki07,mayama10,takakuwa12,dutrey14,pyo14,tobin16,tobin19}. 
%%Observations also indicate that binary formation occurs in single (collapsing) clouds \citep{sadavoy17}.
%%Thus, it is considered that binary systems can form in gravitationally collapsing clouds \citep{schaefer06}. 
It is well known that protostellar jets and outflow are driven by magnetic effects in collapsing clouds \citep{tomisaka02}. 
Recent theoretical studies have shown that the angular momentum in a gravitationally collapsing cloud is effectively transported by magnetic effects such as protostellar jets and magnetic braking \citep[e.g.,][]{vaytet18}.
It is expected that binary formation is closely related to the mechanisms of angular momentum transport \citep{machida08}. 
Thus, we need to carefully consider the effects of the magnetic field when investigating binary systems formed in gravitationally collapsing clouds.
Note that the density perturbation of the initial cloud also affects the (close) binary formation \citep{machida04,machida05b,price07}.

Unlike the single-star formation process, binary formation is very complicated. 
Thus, binary formation and fragmentation process has been investigated in three-dimensional simulations \citep{tsuribe99, matsumoto03,goodwin07, hennebelle08}.
However, only the gas collapse phase before protostar or protobinary formation has been intensively investigated in these studies. 
Also, some studies have ignored the magnetic field, even though the magnetic field significantly affects both the fragmentation process and binary orbital evolution \citep{matsumoto15}. 
Circumbinary outflow and protostellar jets, which are proof of mass accretion onto a binary system, cannot be reproduced without considering the magnetic effects.
\citet{machida09} and \citet{kuruwita17} investigated the formation and evolution of a binary system in three-dimensional magnetohydrodynamics (MHD) simulations and reproduced the circumbinary disk and low-velocity outflow \citep[see also][]{kuruwita19}.
However, they could not reproduce the high-velocity jets, because the protostars were not resolved in their studies. 
High-velocity jets driven by each protostar in a binary system have not yet reproduced in theoretical studies, while observations are revealing high-velocity jets in binary systems \citep{tobin19,hara20}.   
In addition, it is not possible to investigate close binary systems without sufficient spatial resolution in numerical simulations.
With non-ideal MHD simulations, \cite{wurster17}  investigated the formation of binary systems that show neither outflows nor jets, which is attributed to the initial condition adopted in their study (for details, see \S5.5 of \citealt{wurster19}).

%%This study focuses on binary formation and the driving of high-velocity jets. 
%%Starting from a prestellar cloud, we calculate the cloud evolution, the fragmentation process and the evolution of a protobinary system, in which each protostar is spatially resolved.

\section{Numerical Settings and Model} \label{sec:model}
The numerical settings adopted in this study are almost the same as in \citet{machida14} and \citet{machida19}, in which jet driving and disk formation around a single protostar were investigated. 
%%This study  slightly differs from our previous studies in the initial conditions of the prestellar cloud. 
We calculate the cloud evolution from the prestellar stage until $\sim400$\,yr after protostar (or protobinary) formation using  our {\rm resistive} magnetohydrodynamics (MHD) nested grid code, in which equations [1]--[7]  of \citet{machida12} are solved.
An ohmic dissipation term is included in the induction equation, and the coefficient of ohmic resistivity is described in \citet{machida07}.
As described in \citet{machida19}, we adopt the stiff equation of state (EOS) method, which mimics a protostar in a high-density region without sink cells \citep{tomisaka02,hennebelle09,joos12,hirano17}.
The equation of state used in the range of $n<5 \times 10^{16}\cm$ is the same as in \citet{machida14}, while the EOS at high densities of $n>5\times10^{16}\cm$ has a polytropic index $\gamma=2.0$.  
This slight change in the EOS can accelerate the calculation, while slightly expanding the protostar \citep{machida15}. 
%%With this treatment, the protostar has a radius of $\sim0.05$\,au which is about 10 times larger than the solar radius. 
It should be noted that we cannot use sink cells having an accretion radius $\gtrsim1$\,au when investigating high-velocity jets, because the jet driving region is embedded in the sink.

As an initial state, we take a Bonnor--Ebert (BE) sphere with a central density $n_{\rm 0} = 3\times10^6\cm$ and an isothermal temperature $T_{\rm iso,0}=10$\,K. 
The radius of the initial cloud is twice the critical BE radius, which corresponds to $R_{\rm cl}=5.3\times10^3$\,au. 
To promote contraction and realize a gravitationally unstable state, the density of the BE sphere is enhanced by a factor of $f=6.98$ where $f$ is the density enhancement factor \citep[for details, see][]{matsushita17}. 
Thus, the central density of the initial cloud is $n_{\rm c,0}=2.1\times10^7\cm$ ($=f \times n_{\rm 0}$). 
In addition, we added 10\% of the $m=2$ mode of the density perturbation \citep[see][]{machida05}. 
The mass of the initial cloud is $M_{\rm cl}=3.7\msun$. 
A uniform density of $2.3\times10^5\cm$ is set outside the prestellar cloud.  
A uniform magnetic field $B_0=1.2\times10^{-4}$\,G and a rigid rotation $\Omega_0=9.5\times10^{-13}$\,s$^{-1}$ are adopted, in which the magnetic vectors  are parallel to the rotation axis or the $z$-axis. 
The ratio of thermal $\alpha_0$, rotational $\beta_0$ and magnetic $\gamma_0$ energy to the gravitational energy of the prestellar cloud are $\alpha_0=0.1$, $\beta_0=0.05$ and $\gamma_0= 0.03$, respectively. 
The mass-to-flux ratio normalized by the critical value $(2 \pi G^{1/2})^{-1}$ is $\mu_0=5$. 
%%These settings for the initial cloud are almost the same as in our previous studies \citep{machida19}. 

To calculate the cloud evolution and binary formation, we use the nested grid code \citep{machida04}.
Grids having different cell widths are nested and the grid level is described by $l$. 
Each grid is composed of $(i, j, k) = (128, 128, 128)$ cells, and the grid size $L(l)$ and cell width $h(l)$ halve with each increment of the grid level. 
Before the calculation starts, six levels of grid $l=6$ are set for the initial state. 
The initial cloud is immersed in the fourth level of the grid ($l=4$), which has twice the cloud radius $L(4)=2.11\times10^{4}$\,au and a cell width of $h(4)=165$\,au. 
The coarsest grid has a grid size of $L(1)=1.69\times10^5$\,au and a cell width of $h(1)=1332$\,au. 
After the calculation starts, a new finer grid is generated to ensure the Truelove condition, in which the Jeans wavelength is resolved for at least 16 cells. 
The maximum grid level is set to $l=16$ and has $L(16)=5.01$\,au and $h(16)=0.039$\,au.

\section{Results} \label{sec:results}
\begin{figure}[ht!]
\plotone{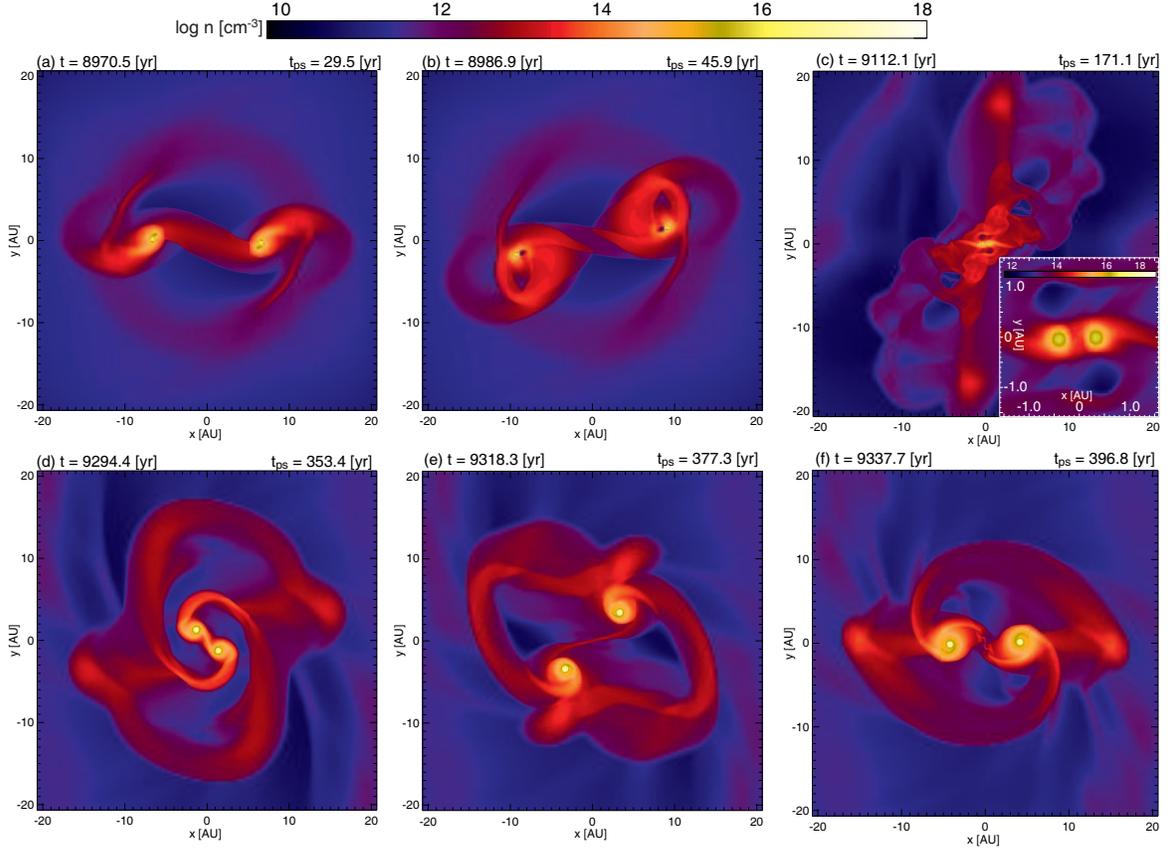}
\caption{
%%Time sequence of a protobinary system.
The density distribution (color) on the equatorial plane is plotted.
The elapsed time after the cloud begins to collapse $t$ and that after protostar formation $t_{\rm ps}$ are given in the upper part of each panel. 
Inset in panel (c) is the close-up view of the central region.
}
\label{f1}
\end{figure}

%%We calculated the cloud evolution from the prestellar stage until $\sim400$\,yr after protostar (or protobinary) formation, though we only show the evolution of a protobinary system in the main accretion phase after protostar formation. 
%%The gas collapse phase before the main accretion phase has already been investigated in our previous studies. 
Figure~\ref{f1} shows the time sequence of a protobinary system for $\sim400$\,yr. 
Note that the structures seen in Figure~\ref{f1} are very similar to those seen in \citet{wurster17}, in which sink particles were used. 
This indicates that the fragmentation and binary formation process can be accurately calculated with sink particles. 
In the collapsing cloud, a bar-like structure develops and fragmentation occurs, as shown in Figure~\ref{f1}{\it a}. 
The high-density region continues to collapse and a protostar forms in each fragment.
Each protostar has a central density of $>10^{15}$--$10^{18}\cm$ and a radius  of $\lesssim0.05$\,au. 
The cavity-like structures or low-density regions in the proximity of the protostars seen in Figure~\ref{f1}{\it b} are caused by magnetic interchange instability, which is usually confirmed in single-star-formation simulations \citep[e.g.,][]{machida20}.  
When fragmentation occurs, the separation between fragments or protostars is about 20\,au (Figs.~\ref{f1}{\it a} and {\it b}). 
As seen in Figure~\ref{f1}({\it a})--({\it c}), the binary separation gradually shrinks during $t_{\rm ps}\lesssim 200$\,yr, where $t_{\rm ps}$ is the elapsed time after protostar formation. 
The protobinary system has a minimum separation $r_{\rm sep}\sim 1$\,au at $t_{\rm ps}\sim170$\,yr (Fig.~\ref{f1}{\it c}).
The separation then increases and maintains $r_{\rm sep}\sim5$--$15$\,au by the end of the simulation (Fig.~\ref{f1}{\it e} and {\it f}). 
In Figure~\ref{f1}{\it d}-{\it f}, we can confirm that each protostar is surrounded by a circumstellar disk with a size of $\sim3-5$\,au. 
In addition, a circumbinary streams encloses the two protostars and their disks with a size of $10-20$\,au\footnote{ 
Here, we call the high-density region that surrounds two protostars the circumbinary streams (red ring-like structure in Figs.~\ref{f1}{\it d}--{\it f}).
%% The rotation of the circumbinary disk can be confirmed in Fig.~\ref{f2}{\it c}.  
}. 
It should be noted that  the binary orbital motion would be related to the amplification of the magnetic field that may depend on the spatial resulution. 
Thus, we need to investigate the spatial resolution necessary to more precisely calculate the binary orbital motion in future studies. 

\begin{figure}[ht!]
\plotone{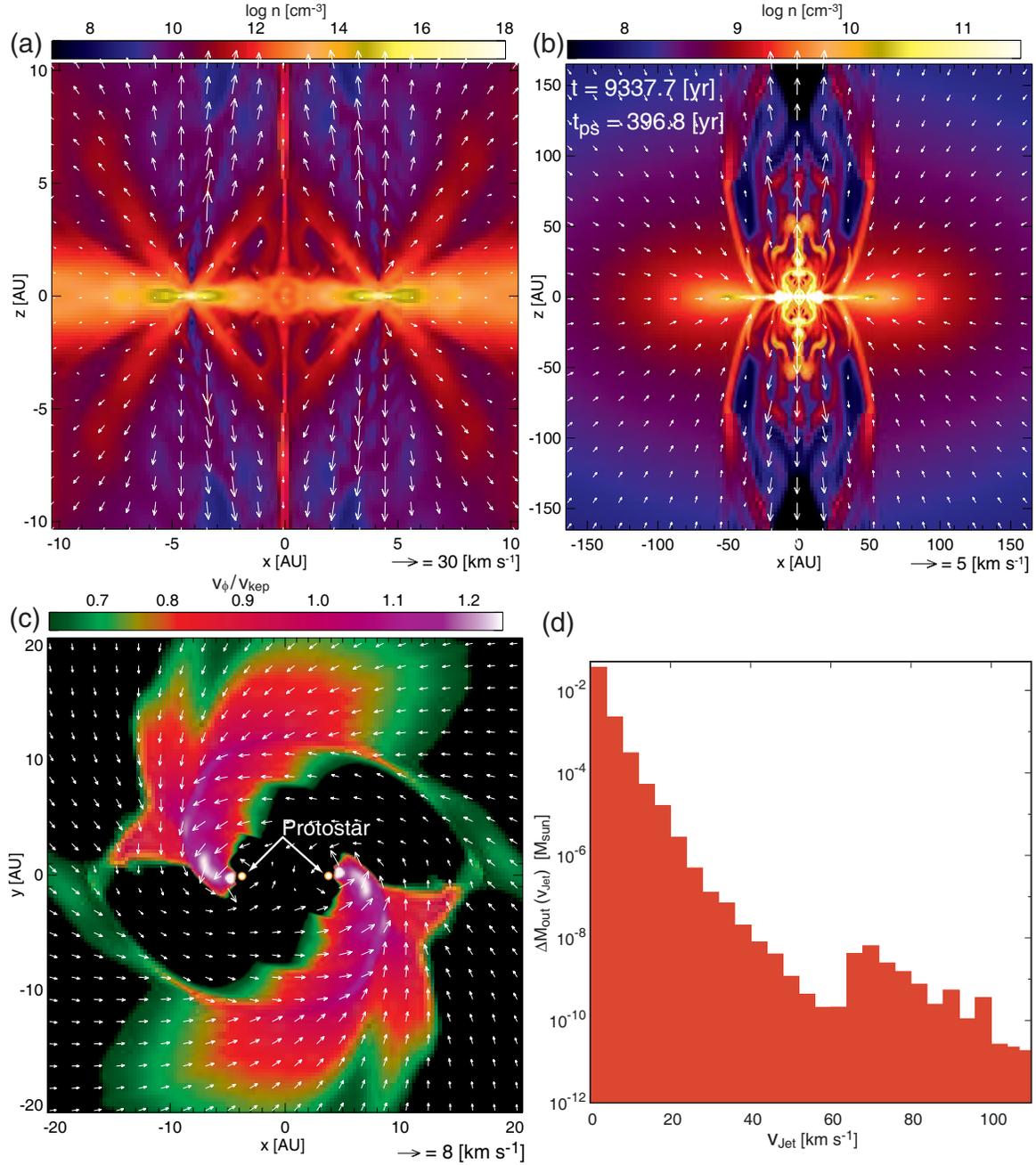}
\caption{
(a), (b) Density (color) and velocity (arrows) distributions on the $y=0$ plane.
Panel (a) is an enlarged view of panel (b). 
(c) Ratio of azimuthal to Keplerian velocity $v_\phi/v_{\rm kep}$ (color) and velocity distribution (arrows) on the equatorial plane. The position of protostars are indicated.  
(d) Histogram of outflowing gas against the outflow velocity. 
The elapsed time after the cloud begins to collapse $t$ and that after protostar formation $t_{\rm ps}$ are given in panel (b).
}
\label{f2}
\end{figure}

Figures~\ref{f2}(a) and (b) show the density distribution at the same epoch as in Figure~\ref{f1}{\it f}.
At this epoch, the protostars are located along the $y=0$ axis as shown in Figure~\ref{f1}{\it f}. 
%%To confirm the jets driven by each protostar, Figure~\ref{f2}{\it a} plots the density and velocity distributions on the $y=0$ plane. 
As seen in Figure~\ref{f2}{\it a}, both protostars drive the high-velocity jets.
In addition, we can confirm cavity-likes structure above and below each protostar. 
On a large scale (Fig.~\ref{f2}{\it b}),  we cannot distinguish each jet because the protostellar jets are highly tangled. 
In addition to the high-velocity jets,  a wide-angle  low-velocity outflow is driven by the circumbinary region as shown in Figure~\ref{f2}{\it b}. 
Note that a low-velocity outflow also appears in the outer region of each circumstellar disk. 
Thus, the outflow at the large scale has an internal structure which is attributed to both the tangled high-velocity jets and low-velocity outflows.
The disk like-structure can be confirmed along the $z=0$ axis in Figure~\ref{f2}{\it b}. 
In addition, outside the protostars, we can confirm the high-density region corresponding to the circumbinary streams with a radius of $\sim10-15$\,au in Figure~\ref{f1}{\it d}-{\it f}.
Figure~\ref{f2}{\it c} shows the ratio of azimuthal  to Keplerian velocity ($v_\phi/v_{\rm kep}$), in which the central mass $M_{\rm c}$ is derived by the sum of the gas in the region with $n>10^{13}\cm$ in the range of $<10$\,au to estimate  the Keplerian velocity  $v_{\rm kep}=(G M_{\rm c}/r)$.
Note that the mass within the circumbinary streams are concentrated within $<10$\,au (Figs.~\ref{f1}{\it d}-{\it f}). 
The figure indicates that the rotation velocity is comparable to the Keplerian velocity in a large part of the circumbinary streams. 
%%The gravitational torque induced by the binary orbital motion should effectively transfer the angular momentum of the circumbinary disk.
%%Thus, the morphology of the circumstellar disk changes with time  (see Fig.~\ref{f1} and movie 1). 
Thus, it is natural that the circumbinary streams  can drive the (low-velocity) outflow as seen in \citet{machida09} and \citet{kuruwita17}.
The outgoing flow reaches $\sim 200$\,au from the protostars by the end of the simulation.

Figure~\ref{f2}{\it d} plots the outflowing mass $\Delta M_{\rm out}$ in different outflow velocity bins ($\Delta v_r=4\km$) at $t_{\rm ps}=396.8$\,yr and shows that the protobinary system drives the outflow mainly in the range of $\lesssim 100 \km$ at this epoch, in which  a small fraction of outgoing flow  exceeds $100\km$.
The figure also indicates that the low-velocity component dominates the high-velocity component.
The low-velocity component is mainly driven by the circumbinary region, while the high-velocity component appears near the protostars. The local maximum around $v_r\sim70\km$ corresponds to the high-velocity jets directly driven by each protostar. 

\begin{figure}[ht!]
\plotone{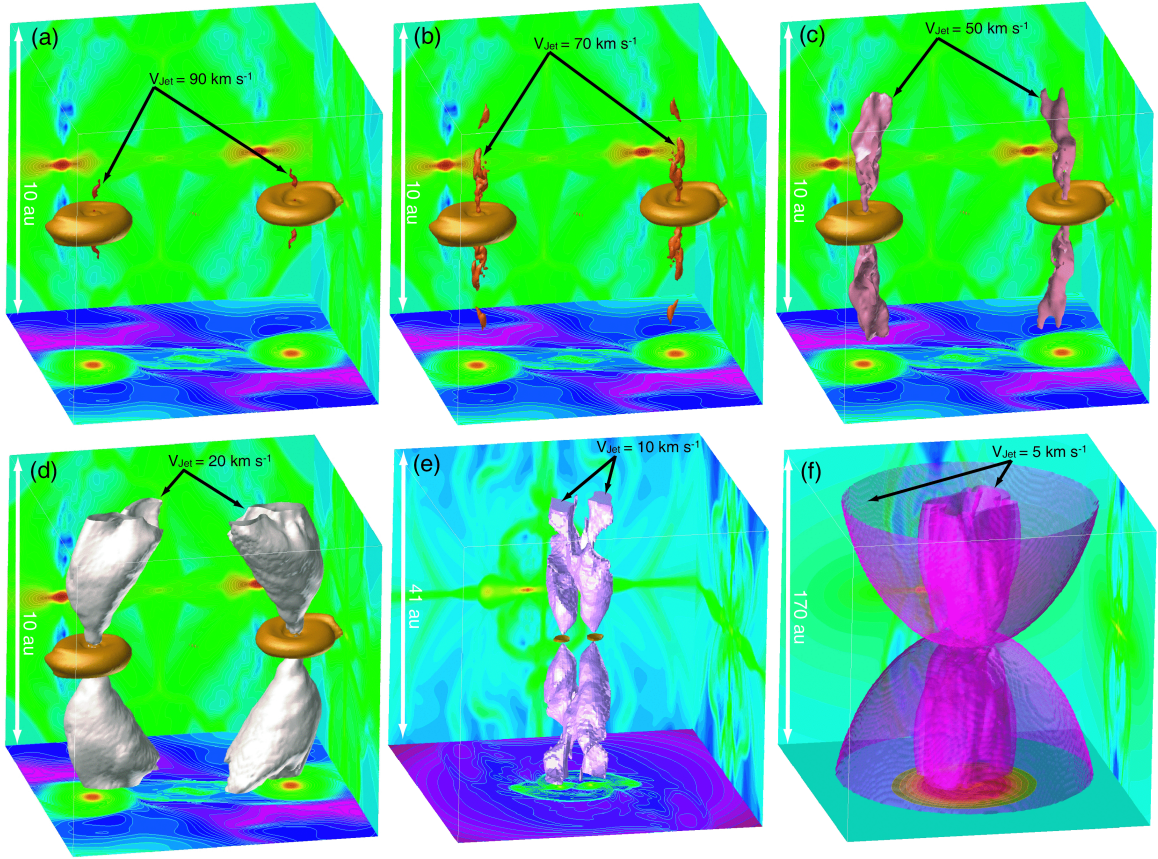}
\caption{
The structure of high-velocity jets at the same epoch as in Fig.~\ref{f2} with an iso-velocity surface of (a) $v_{\rm Jet}=90\km$, (b) $70\km$, (c) $50\km$, (d) $20\km$ (e) $10\km$ and (f) $5\km$.  
The yellow surface corresponds to the circumstellar disk with an iso-density surface of $n=10^{13} \cm$ within which a protostar is embedded. 
The density distribution on the $x=0$, $y=0$ and $z=0$ cutting planet are projected on each wall surface. 
The box scale is described in each panel. 
}
\label{f3}
\end{figure}

Figure~\ref{f3} shows the three-dimensional structures of the high-velocity jets at  $t_{\rm ps}=396.8$\,yr, in which the jet structures with different velocity components are delineated. 
At this epoch, a very small amount of the high-velocity components ($>90\km$) appears above and below the protostars and the jets are considerably distorted (Fig.~\ref{f3}{\it a}), while the velocity components of $70\km$ are directly connected to the protostars (Fig.~\ref{f3}{\it b}). 
The jet velocity of $v_{\rm Jet} \simeq 70\km$ roughly corresponds to the Keplerian velocity just outside the protostar where the Keplerian velocity  $(GM_{\rm ps}/r_{\rm ps})^{1/2}$ becomes  $57-67\km$ with a protostellar mass  of $M_{\rm ps} \simeq 0.15\msun$ (see below) and a protostellar radius of $r_{\rm ps} \simeq 0.03-0.04$\,au (Fig.~\ref{f2}{\it c}). 
Note that a small fraction of the jet driven by each circumstellar disk and protostar can be further accelerated near the driving region \citep[e.g.,][]{kudoh98} and produce the very high-velocity components ($>70\km$).
The structures of the jet with $\gtrsim 50\km$ have a  well collimated structure (Figs.~\ref{f3}{\it b} and {\it c}), while the collimation of relatively low-velocity components of $\lesssim 20\km$ is not very good. 
The jets are tangling on a scale of $\sim40$\,au (Fig.~\ref{f3}{\it e}), while the highly tangled jets are confirmed like a single distorted jet on a large scale (Fig.~\ref{f3}{\it f}).
Note that, in Figure~\ref{f3}{\it f}, the cone-line structure enclosing the central jet corresponds to the outflow driven from the circumbinary region.   
The figure indicates that the spatial resoluton of $\lesssim 100$\,au is required to resolve each jet driven by each protostar of close binary system in observetion.

\begin{figure}[ht!]
\epsscale{0.85}
\plotone{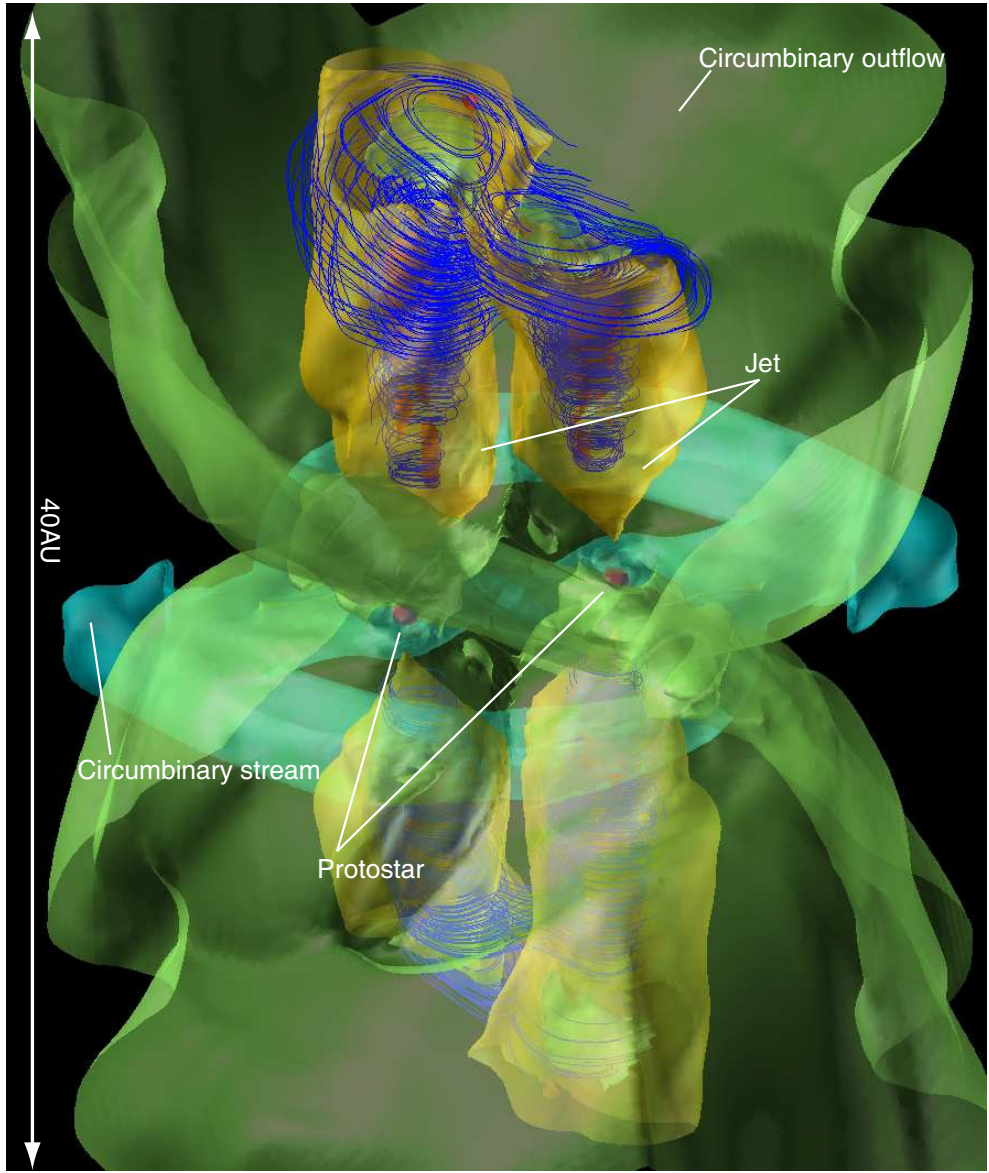}
\caption{
Three-dimensional view of protostars (red iso-density surfaces, $n=10^{16}\cm$), circumstellar disks, circumbinary stream (blue iso-density surfaces, $n=10^{13}\cm$), high-velocity jets (yellow iso-velocity surfaces, $v_r=30\km$) and circumbinary outflow (green iso-velocity surfaces, $v_r=5\km$).
The blue lines are magnetic field lines. 
The box size is 40\,au. 
}
\label{f4}
\end{figure}

Figure~\ref{f4} shows the three-dimensional structure of a protobinary system at the same epoch as in Figure~\ref{f3}.
We can clearly confirm twin jets driven by protostars.
Since each protostar orbits in a counterclockwise fashion, the protostellar jets are somewhat swung back in the same direction. 
Within the jets, magnetic field lines are strongly twisted. 
Near the roots of the jets, the protostars are enclosed by the circumstellar disks, which are surrounded by the circumbinary streams (see also Figs.~\ref{f1} and \ref{f2}).
%%In the simulation, since the $m=2$ mode of the density perturbation is added to the initial state, the two protostars have almost the same mass and maximum density by the end of the simulation. 
%%Thus, we do not distinguish the two protostars, such as a primary and a secondary star, in this paper. 
%%As seen in Figure~\ref{f1}, the orbital motion of the protostars is not simple. 
%%The binary separation should be determined by the balance between the influx and outflux of the angular momentum. 
%%Since the accreting gas has angular momentum, the binary system receives angular momentum from the infalling gas. 
%%On the other hand, the angular momentum of the binary system is transported by magnetic, thermal pressure and gravitational torques.
%%Especially, magnetic effects such as magnetic braking and protostellar jets and outflows would play an important role for transporting the angular momentum  \citep{machida19}.  
%%It is very difficult to identify which torque is most dominant because the system is very complicated. 
%%We will focus on the mechanism of angular momentum transport in binary systems in a future study.   

\begin{figure}[ht!]
\plotone{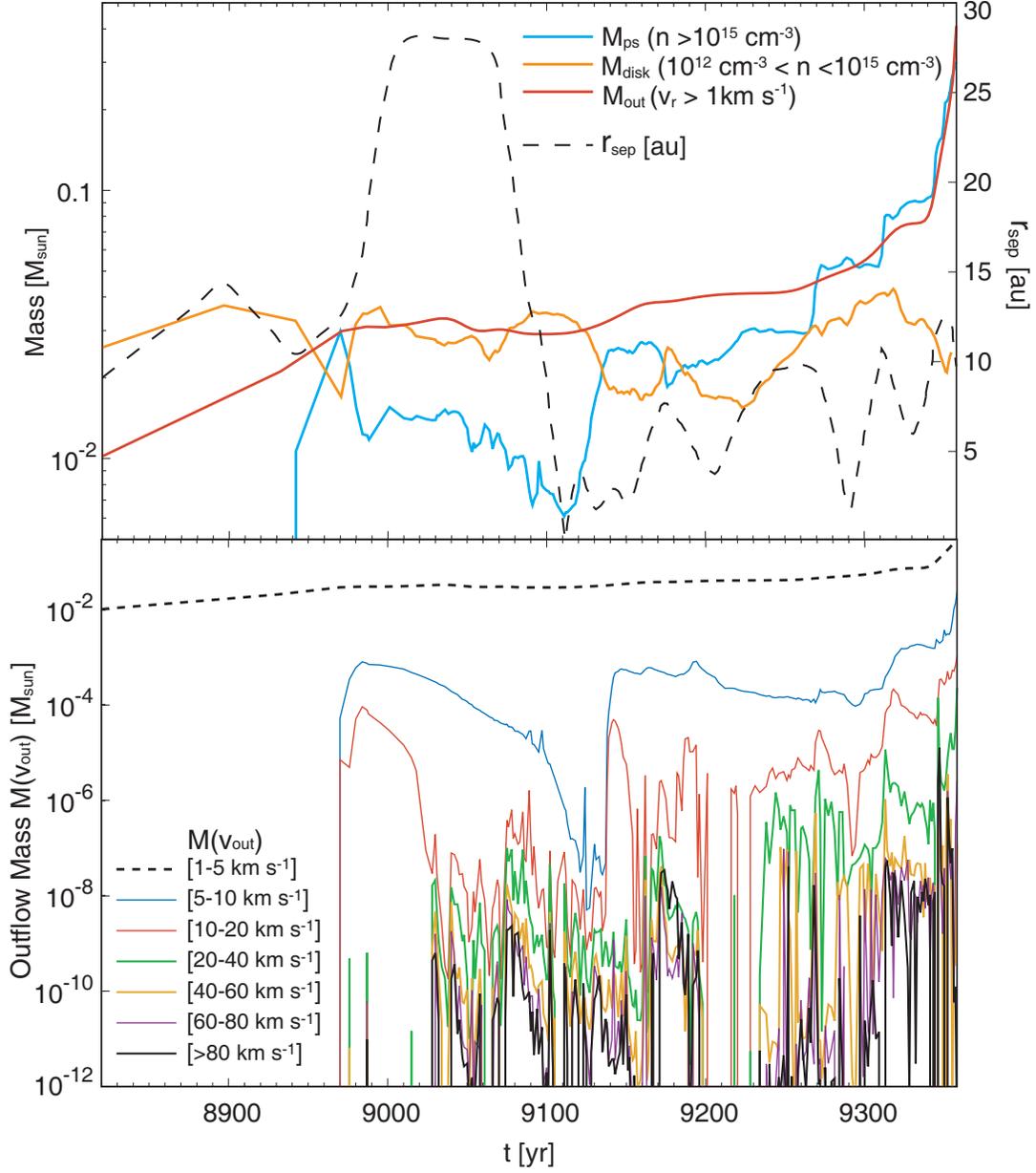}
\caption{
Protostellar, disk and outflow mass against the elapsed time $t$ (top; left vertical axis).
Outflow mass in different velocity range against the elapsed time (bottom; left vertical axis).
The binary separation $r_{\rm sep} $(right axis) is plotted in the top panel.
}
\label{f5}
\end{figure}

Figure~\ref{f5} top panel shows the mass of the high-density regions and outflow.
%%Since we did not use sink cells, it is difficult to distinguish the boundary between the protostar and the circumstellar disk or circumbinary disk. 
Roughly, we estimated the mass of the high-density region with $n>10^{15}\cm$ as being the protostar and that with $10^{12}\cm < n < 10^{15}\cm$ to be the circumstellar disk, which are shown in Figure~\ref{f5} top panel. 
%%We determined the density range of each object by checking the density distribution of the protobinary system as seen in Figure~\ref{f1}.
The figure indicates that the high-density region ($n>10^{15}\cm$) appears at $t=8941$\,yr, which corresponds to the protostar formation epoch ($t_{\rm ps}= 0$). 
Note that a temporal decrease of the mass of the high-density region ($n>10^{15}\cm$) is due to the oscillation of the high-density objects (i.e., protostars). 
The protostar bounces and its (average) density slightly decreases after the protostar shrinks with a temporal high mass accretion rate. 
The mass with $n>10^{15}\cm$ (total mass of protostars) reaches $\simeq 0.3\msun$ at the end of the simulation, while that with $10^{12}\cm < n < 10^{15}\cm$  (total mass of circumstellar disks) is $\sim0.03\msun$ during the simulation.
Since there are two protostars and circumstellar disks, the mass of the protostar and circumstellar disk are estimated to be $M_{\rm ps}\simeq0.15\msun$ and $M_{\rm disk}\simeq0.015\msun$, respectively.
On the other hand, the outflow mass, which is defined as the total mass having $v_r > 1\km$, increases from $M_{\rm out}\simeq0.01\msun$ to $0.4\msun$ during the simulation, and is comparable to the protostellar mass at the end of the simulation.
Thus, a significant mass ejection is realized in the protobinary system, as seen in the single-star formation process \citep[e.g.,][]{machida19}.

The binary separation is also plotted in the top panel of Figure~\ref{f5}. 
%%As seen in Figure~\ref{f1}, the binary orbital motion is not simple.
The binary separation oscillates in the range of $1 \,{\rm au} \lesssim r_{\rm sep} \lesssim 30\,{\rm au}$. 
We can see a rough correlation between the masses of the protostar, disk and outflow  and  the binary separation. 
To investigate the relationship between the binary separation and the outflow driving, the outflowing mass with different velocity ranges are plotted in Figure~\ref{f5} bottom panel. 
The figure indicates that, during the early main accretion phase, the outflow mass of the low-velocity component ($v_r=1-5\km$) dominates that of the high-velocity component ($v_r>5\km$, see also Fig.~\ref{f3}{\it d}) and the low-velocity component is not significantly affected by the binary orbital motion. 
In addition, the low-velocity component  ($v_r<5\km$) appears before protostar formation $t_{\rm ps}<0$ (or $t<8941$\,yr), while the high-velocity component ($v_r>5\km$) appears $t_{\rm ps}\gtrsim30$\,yr (or $t\gtrsim8970$\,yr) after protostar formation.  
The low-velocity outflow is originally driven by the first core \citep{wurster18}, which forms before protostar formation and evolves into the circumbinary structure. 
On the other hand, the high-velocity components are driven near the protostar where the gravitational potential has a local minimum (Fig.~\ref{f3}).
Thus, the difference in the flow emergence epochs is due to the different formation epochs of each object. 
It should be noted that the low-velocity flow appears before the high-velocity flow even in the single star formation simulations \citep{machida19}. 
Thus, the different emergence epochs of the flows is not a unique feature for the binary or multiple star formation process but universally occurs in the star formation process. 
In Figure~\ref{f5} bottom panel, the time variability in the outflow mass is more significant in relatively high-velocity components ($v_r>10$--$20\km$) than in relatively low-velocity components ($v_r<10\,\km$). 
Especially, the outflow mass in the very high-velocity components ($v_r>20\km$) seems to correlate with the binary orbital separation, in which a strong mass ejection occurs when the binary separation becomes large.
%%Since the binary protostars orbit each other only for several times during the simulation, we need a further time-integration to clarify the relationship between the jet activity and the binary separation. 

\section{Summary and Discussion} \label{sec:summary}
We could reproduce a protobinary system in a core collapse simulation and confirmed the presence of  protostellar jets and circumbinary outflow, which are usually observed in very young binary systems. 
Especially, for the first time, we could reproduce high-velocity jets with a maximum speed of $\gtrsim 100\km$ driven by each protostar in the protobinary system, bridging the gap between theoretical studies and observations. 
The high-velocity components (or protostellar jets) show a significant time-variability, while the low-velocity components do not show a noticeable time-variability (Fig.~\ref{f5}).
In this section, we roughly estimate the necessary spatial resolution to observe the jets driven from protobinary systems.
Although we conceived and referred to the simulation results, the following quantitative estimates are not directly related to quantities in the  simulation. 
The typical timescale of the system should  be determined by the binary orbital period $P=(4\pi^2 a^3/G M_{\rm tot})^{1/2}$, where $a$ and $M_{\rm tot}$ are the binary orbital radius and total mass of binary protostars, respectively.
Simply assuming the binary orbital radius of 10\,au and total mass of $0.1\msun$, the orbital period $P=100 (a/10\,{\rm au})^{3/2} (M_{\rm tot}/0.1\msun)^{-1/2}$\,yr is derived.
We also assume that the circumbinary disk (or stream) has a radius of $>10$\,au within which protostars are embedded. 
In such a case, the orbital period of the outer circumstellar disk is $>100$\,yr.
Since the low-velocity outflow is driven by the circumbinary disk (or stream)  which is located far from the protostars (Fig.~\ref{f5} bottom panel), it is not significantly disturbed by the binary orbital motion. 
On the other hand, the jets are easily disturbed because the jet launching points orbit with a period of $<100$\,yr in close binary systems.
Further, assuming a typical jet velocity of $30\km$, the jets reach $L_{\rm Jet}\sim 600\, (v_{\rm Jet}/30\km)$\,au during one orbital period of 100\,yr.
Note that the velocity of $30\km$ roughly corresponds to the typical jet velocity in the simulation. 
Thus, only jets with a size of $ \ll L_{\rm Jet}$ are detectable, while a complex outflow would be detected in the range  of $\gtrsim L_{\rm Jet}$. 
We can actually distinguish the binary jets within $<100$\,au in the simulation (Figs.~\ref{f3} and \ref{f4}). 
Resolving two protostellar jets in observations is evidence of the existence of a protobinary system. 
Our study showed that a close binary system with a separation of $\sim10$\,au can drive twin jets with a size of $\sim10-100$\,au, which is observable by current telescopes.

\section*{Acknowledgements}
We thank the referee for very useful comments and suggestions on this paper. 
The present research used the computational resources of the HPCI system provided by the Cyber Science Center, Tohoku University and Cybermedia Center, Osaka University, Earth Simulator, JAMSTEC, through the HPCI System Research Project (Project ID:hp180001,hp190035,hp200004).
The present study was supported by JSPS KAKENHI Grant Numbers JP17K05387, JP17H02869,  JP17H06360 and  17KK0096.
The simulations reported in this paper were also performed by 2017 and 2018 Koubo Kadai on the Earth Simulator (NEC SX-ACE) at JAMSTEC.

\end{document}